# Visualizing the Odd-parity Superconducting Order Parameter and its Quasiparticle Surface Band in UTe$_2$

Shuqiu Wang[1,2] and J.C. Séamus Davis[2,3]
1. H. H. Wills Physics Laboratory, University of Bristol, Bristol, BS8 1TL, UK
2. Clarendon Laboratory, University of Oxford, Oxford, OX1 3PU, UK
3. Department of Physics, University College Cork, Cork T12 R5C, IE

**ABSTRACT**

A distinctive identifier of nodal intrinsic topological superconductivity (ITS) would the appearance of an Andreev bound state on crystal surfaces parallel to the nodal axis, in the form of a topological quasiparticle surface band (QSB) appearing only for $T < T_C$. Moreover, theory shows that specific QSB characteristics observable in tunneling to an *s*-wave superconductor can distinguish between chiral and non-chiral ITS order parameter $\Delta_k$. To search for such phenomena in UTe$_2$, *s*-wave superconductive scan-tip scanning tunneling microscopy (STM) imaging was employed. It reveals an intense zero-energy Andreev conductance maximum at the UTe$_2$ (0-11) crystal termination. Development of the zero-energy Andreev conductance peak into two finite-energy particle-hole symmetric conductance maxima as the tunnel barrier is reduced, then signifies that UTe$_2$ superconductivity is non-chiral. Quasiparticle interference imaging (QPI) for an ITS material should be dominated by the QSB for energies within the superconductive energy gap $|E| \leq \Delta$, so that bulk $\Delta(k)$ characteristics of the ITS can only be detected excursively. Again using a superconducting scan-tip, the in-gap quasiparticle interference patterns of the QSB of UTe$_2$ were visualized. Specifically, a band of Bogoliubov quasiparticles appears as a characteristic sextet $q_i: i = 1 - 6$ of interference wavevectors showing that QSB dispersions $k(E)$ occur only for energies $|E| \leq \Delta_{max}$ and only within the range of Fermi momenta projected onto the (0-11) crystal surface. In combination, these phenomena are consistent with a bulk $\Delta(k)$ exhibiting spin triplet, time-reversal conserving, odd-parity, *a*-axis nodal, *B$_{3u}$* symmetry in UTe$_2$.

**Keywords**  Intrinsic Topological Superconductivity. Andreev and Josephson Scanning Tunneling Microscopy.

**Intrinsic Topological Superconductivity**

For spin-triplet superconductors [1-4], the order parameter $\Delta_k = \begin{pmatrix} \Delta_{k\uparrow\uparrow} & \Delta_{k\uparrow\downarrow} \\ \Delta_{k\downarrow\uparrow} & \Delta_{k\downarrow\downarrow} \end{pmatrix}$ with $\Delta^T_{-k} = -\Delta_k$ and $\Delta_k = \Delta^T_k$, is also represented in the ***d***-vector notation as $\Delta_k \equiv \Delta_0(\boldsymbol{d} \cdot \boldsymbol{\sigma})i\sigma_2$ where $\sigma_i$ are the Pauli matrices. In principle, such systems are ITS whose signature is the existence of an odd-parity bulk superconducting energy gap, along with the presence of symmetry-protected gapless topological quasiparticle surface bands of Bogoliubov quasiparticles within that gap. When such superconductors are topological [5], it is not because of electronic band-structure topology but because $\Delta_k$ itself exhibits topologically non-trivial properties [6]. The search for technologically viable ITS is now a forefront of quantum matter research [7]. Here we summarize recent scanning tunneling microscopy studies, specifically using superconducting scan-tips in the Josephson and Andreev modes as explained below, of the $\Delta_k$ and associated quasiparticle surface bands in the candidate ITS material, UTe$_2$.

**Prevenient Charge Density Wave State in UTe$_2$**

Well above the superconducting critical temperate $T_C$, three charge density wave (CDW) states with distinct wavevectors are observed [8] at the equivalent (0-11) cleave surface of UTe$_2$ where our studies are carried out. These states have not been detected in bulk [9,10]. Upon entering the superconductive state, three pair density wave (PDW) states with distinct wavevectors are observed, through their periodically modulating superconducting energy gap, at the identical wavevectors as the prevenient CDWs [11]. These phenomena are all consistent with induction of these UTe$_2$ PDW states due to the interactions between the prevenient CDW states and the superconductivity. At present, these UTe$_2$ PDW states have not played a role in determination of the symmetry of the bulk superconductor order parameter.

**Superconducting Order Parameter of UTe$_2$**

The recently discovered superconductor UTe$_2$ is the leading candidate to be a 3D nodal spin-triplet superconductor [12,13] and thus an ITS. The crystal symmetry point-group is $D_{2h}$ so that there are four possible odd-parity order parameter symmetries designated $A_u$, $B_{1u}$, $B_{2u}$ and $B_{3u}$. All of these preserve time reversal symmetry: $A_u$ is fully gapped whereas $B_{1u}$, $B_{2u}$ and $B_{3u}$ have nodes in $\Delta_k$, whose axial alignment is along lattice vectors ***c***, ***b*** or ***a*** respectively. If they are accidentally degenerate, linear combinations of these order parameters are also possible which break point-group and time reversal symmetries, resulting in a chiral QSB along with persistent surface supercurrents orthogonal to the nodal-axis [6,7]. For UTe$_2$, there are two chiral states of particular interest with $\Delta_k$ nodes aligned with the crystal ***c***-axis, and two with nodes aligned with the ***a***-axis. Identifying which (if any) of these $\Delta_k$ exists in UTe$_2$ is key to demonstrating and utilizing the novel physics of this material.

However, this has proven a challenging objective [14]. For example, a magnetic susceptibility upon entering the superconducting phase that is equivalent to Pauli paramagnetism, is deduced from minimal suppressions of the Knight shift [15] and used to adduce spin-triplet pairing. Some NMR studies measuring the change of the spin susceptibility across $T_c$ report a decrease in the Knight shift in all directions and hypothesize the $A_u$ state [15], whereas other NMR studies detect a reduction in the Knight shift along the ***b*** and ***c*** axes only, thence hypothesizing $B_{3u}$ state [16]. Magnetic field orientation of the thermal conductivity indicates point nodes parallel to the crystal ***a***-axis [17], whereas other field-oriented thermal conductivity measurements [18] report isotropic results and hypothesize an $A_u$ symmetry. Field-oriented specific heat measurements reveal peaks around the crystal ***a***-axis implying point nodes oriented along this direction and hypothesize an order parameter with chiral $A_u + iB_{3u}$ or helical $B_{3u}$ symmetries [19]. Some electronic specific heat studies report two specific heat peaks and hypothesize a chiral $A_u + iB_{1u}$ or $B_{2u}$

+ $iB_{3u}$ order parameter [20], whereas other specific heat studies detect only a single specific heat peak and thus hypothesize a single component order parameter [21]. London penetration depth measurements of superfluid density report anisotropic saturation consistent with nodes along the *a*-axis suggesting $B_{3u}$ symmetry pairing for a cylindrical Fermi surface [22], while other penetration depth measurements exhibiting an $n \leq 2$ power law dependence of the penetration depth on temperature motivate a hypothesis of $B_{3u} + iA_u$ pairing symmetry [23]. Scanning tunneling microscopy experiments in the (0-11) plane parallel to *a*-axis show energy-reversed particle-hole symmetry breaking of some electronic structure elements at opposite UTe2 step edges [24] with the consequent hypothesis of a chiral surface state $B_{1u} + iB_{2u}$ whose nodes are aligned to the *a*-axis. Polar Kerr effect measurements report a field-induced Kerr rotation indicating the presence of time-reversal symmetry breaking and hypothesize chiral $B_{2u} + iB_{3u}$ or $A_u + iB_{1u}$ pairing [20] with nodes aligned to the *c*-axis, whereas other polar Kerr effect measurements report no detectable spontaneous Kerr rotation [25]. However, until recently, no tunneling spectroscopic measurements of $\Delta_\mathbf{k}$, which could differentiate directly between these scenarios, had been reported.

**Scanned Andreev Tunneling Microscopy**

Ideally, $\Delta(\mathbf{k})$ of UTe2 might be established by using Bogoliubov quasiparticle interference imaging, a recognized technique for $\Delta(\mathbf{k})$ determination in complex superconductors [26 - 34]. However, odd-parity superconductors should support a topological [28, 35] QSB on crystal termination surfaces only for energies within the superconductive energy gap $|E| \leq \Delta$. Classically, odd-parity superconductors exhibit zero-energy surface Andreev bound states [36-40] which are generated by the universal $\pi$-phase-shift during Andreev reflections from the odd-parity pair potential $\Delta_k$. More intriguingly, ITS [5,41] exists most simply in the case of odd-parity spin-triplet superconductors. Thus a definitive characteristic [28] of an ITS would be a topological quasiparticle surface band with momentum-energy relationship $\mathbf{k}(E)$ existing only for $T < T_c$, and for energies $|E| \leq \Delta$ within the maximum superconducting energy gap [28,42-52].

QPI is a tunneling spectroscopic technique for establishing $\Delta_k$ in unconventional superconductors. However, normal-tip QPI had also proven ineffective for determining $\Delta_k$ of UTe2 because even at $T/T_c \lesssim 1/7$ a typical quasiparticle density of states spectrum $N(E \leq \Delta_0)$ is essentially metallic with only tenuous hints of opening the bulk $\Delta_k$ [8,11]. The classic QPI signature [26] of scattering interference between $\mathbf{k}$-space locations defined by a bulk superconductive $\Delta_k$ had proven impossible to detect, apparently because the extraordinarily high $N(E \leq \Delta_0)$ of the QSB overwhelms any tunneling conductance signal from the 3D quasiparticles. Hence, the possibility of using a superconductive-tip [53-58] to carry out tunneling spectroscopy in the Andreev mode was explored. There are two channels for conduction from the fully gapped *s*-wave superconductive tip to a nodal spin-triplet superconductor: (a) single-electron tunneling for which the minimum voltage bias is $|V| > \Delta_{\text{tip}}/e$ due to the energy cost $\Delta_{\text{tip}}$ of creating an unpaired electron in the superconducting tip; (b) Andreev reflection of pairs of sub-gap quasiparticles allowing the transfer of $2e$ across the junction, thus generating strong conductance at $|V| < \Delta_{\text{tip}}/e$. Hence, in principle, there are strong advantages to using scanned Andreev tunneling spectroscopy for ITS studies, especially that QSB quasiparticles at the interface between sample and tip predominate the Andreev process (as shown schematically in Fig. 1), and that the order parameter symmetry difference between sample and tip does not preclude the resulting zero-bias Andreev conductance.

Generally, in superconductive-tip scanned Josephson tunneling microscopy, the electron-pair density in a superconductor, $\rho_P(\mathbf{r})$, is visualized by measuring Josephson critical-current $I_J(\mathbf{r})$ from a superconducting STM tip [59], since $\rho_P(\mathbf{r}) \propto I_J^2(\mathbf{r}) R_N^2(\mathbf{r})$ where $R_N$ is the normal-state junction resistance [60,61]. However, thermal fluctuation energy $k_B T$ typically greatly exceeds the Josephson energy $E_J$ so that the tip-sample Josephson junction exhibits a phase-diffusive [ 62 - 64 ] steady-state electron-pair current $I_P(V) = \frac{1}{2} I_J^2 Z V/(V^2 + V_c^2)$ at voltage $V$, where $V_c = 2eZk_B T/\hbar$ and $Z$ is the high-frequency junction impedance. In this case, $dI_P/dV \equiv g(V) = \frac{1}{2} I_J^2 Z (V_c^2 - V^2)/(V^2 + V_c^2)^2$ so that $g(0) \propto I_J^2$. The key consequence is that spatially resolved measurements of $g(\mathbf{r}, 0)$ using superconductive-tip STM at sub-kelvin temperatures now provide a practical technique to visualize electron-pair density $\rho_P(\mathbf{r}) \propto g(\mathbf{r}, 0) R_N^2(\mathbf{r})$ at atomic-scale. Technically closely related is scanned Andreev tunneling microscopy (SATM) [65] which, in theory, is highly advantageous for studying ITS. SATM measures the differential Andreev conductance $a(\mathbf{r}, V) = dI/dV(\mathbf{r}, V)$ and, in the case of ITS, exhibits unique phenomena due to the fact that tunneling occurs from an *s*-wave scan-tip to a *p*-wave ITS through its QSB (Fig. 1).

**Modeling Andreev Tunneling Spectroscopy for ITS**

Novel models are required to understand SATM from an even-parity $\Delta_{\mathbf{k}}$ (e.g. *s*-wave) superconducting scan tip to an odd-parity $\Delta_{\mathbf{k}}$ (e.g. *p*-wave) ITS sample. Most simply, a nodal spin-triplet *p*-wave superconductor on a spherical Fermi surface within a cubic 3D Brillouin zone (BZ) exhibits two nodal points at $\pm \mathbf{k}_n$. Its Hamiltonian is:

$$H = \sum_{k_x} \sum_{\mathbf{k}_\perp} \psi^+(k_x, \mathbf{k}_\perp) h(k_x, \mathbf{k}_\perp) \psi(k_x, \mathbf{k}_\perp). \qquad (1)$$

where $\psi^T(\mathbf{k}) = (c_{\mathbf{k}\uparrow}, c_{\mathbf{k}\downarrow}, c^+_{-\mathbf{k}\uparrow}, c^+_{-\mathbf{k}\downarrow})$ is the Nambu fermion operator, and $h(k_x, \mathbf{k}_\perp)$ is a $4 \times 4$ matrix, containing the information on both band structure and $\Delta_{\mathbf{k}}$ [65]. Considering only a particular 2D slice of the 3D Brillouin zone with a fixed $k_x$, its Hamiltonian $h(k_x, \mathbf{k}_\perp)$ is that of a 2D superconductor within a 2D Brillouin zone spanned by $\mathbf{k}_\perp$. The 2D states $|k_x| < |k_n|$ are topological and those $|k_x| > |k_n|$ are non-topological. The essential signature of such physics is the presence of a QSB also termed an Andreev bound state [6], on the edges of each 2D slice for $|k_x| < |k_n|$. The 2D Brillouin zone of any crystal surface parallel to the nodal axis of $\Delta_{\mathbf{k}}$ has a line of zero-energy QSB states, the so-called Bogoliubov-Fermi Arc, should in theory connect the two points representing the projections of the nodal wavevectors $\pm k_n$ onto this 2D zone. Calculation of the density of QSB quasiparticle states versus energy, $N(E)$ from the QSB dispersion $\mathbf{k}(E)$ yields a continuum in the range $-\Delta_0 \leq E \leq \Delta_0$, with a sharp central peak at $E = 0$ due to the Bogoliubov-Fermi arc. In this picture, the presence or absence of a gapless QSB on a given surface of a 3D crystal, a zero-energy peak in $N(E)$ from the QSB Fermi-arcs, and the response of the QSB to breaking specific symmetries, can reveal the symmetry of the 3D $\Delta_{\mathbf{k}}$.

Considering an *s*-wave superconducting tip (Nb), and a nodal *p*-wave superconductor (UTe2) which sustains a QSB within the interface and are connected by tunneling (SIP model). The Hamiltonian of the SIP model has three elements: $H = H_{Nb} + H_{UTe_2} + H_T$. Here $H_{Nb}$ is the Hamiltonian for an ordinary *s*-wave superconductor given by $H_{Nb}(\mathbf{k}) = \begin{pmatrix} \epsilon_{Nb}(\mathbf{k})\sigma_0 & \Delta_{Nb}(i\sigma_2) \\ \Delta_{Nb}^*(-i\sigma_2) & -\epsilon_{Nb}(-\mathbf{k})\sigma_0 \end{pmatrix}$. Here $\epsilon_{Nb}(\mathbf{k})$ is the band structure model for Nb and $\Delta_{Nb}$ is the Nb superconducting order parameter and $\sigma_{0,1,2,3}$ are the four components of Pauli matrices. $H_{UTe_2}$ is the Hamiltonian of the putative *p*-wave superconductor with $\begin{pmatrix} \epsilon_{UTe_2}(\mathbf{k})\sigma_0 & \Delta_{UTe_2}(\mathbf{k}) \\ \Delta_{UTe_2}^+(\mathbf{k}) & -\epsilon_{UTe_2}(-\mathbf{k})\sigma_0 \end{pmatrix}$. Here $\epsilon_{UTe_2}(\mathbf{k})$ is the band structure model containing the relevant Fermi surface, and $\Delta_{UTe_2}(\mathbf{k})$ is a 2 × 2 spin-triplet pairing matrix given by $\Delta_{UTe_2}(\mathbf{k}) \equiv \Delta_{UTe_2} i(\mathbf{d} \cdot \boldsymbol{\sigma})\sigma_2$. $H_T$ is the tunneling Hamiltonian between the two superconductors $H_T = -|M|\sum_{\mathbf{k}_\parallel}[\psi_{Nb,\mathbf{k}_\parallel}^* \sigma_3 \otimes \sigma_0 \psi_{UTe_2,\mathbf{k}_\parallel}(\mathbf{k}) + h.c.]$ ; $\mathbf{k}_\parallel$ is the momentum in the plane parallel to the interface, $\psi$ is the four-component fermion field localizing on the adjacent planes of Nb and UTe2, and $|M|$ is the tunneling matrix element. To simplify calculation, $\epsilon_{Nb}(\mathbf{k})$ and $\epsilon_{UTe_2}(\mathbf{k})$ are approximated as single bands via a nearest neighbour tight-binding dispersion.

For $H_{UTe_2}$ two scenarios were then considered: (1) chiral pairing state *Au* + *iB3u* with $\mathbf{d}(\mathbf{k}) = (0, k_y + ik_z, ik_y + k_z)$ and, (2) non-chiral pairing state *B3u* with $\mathbf{d}(\mathbf{k}) = (0, k_z, k_y)$. In both examples the two nodes of $\Delta_\mathbf{k}$ lie along the *a*-axis as in Fig. 1, and $\Delta_{UTe_2} = \frac{1}{5}\Delta_{Nb}$ approximate the ratio of maximum energy-gaps of Nb and UTe2. First, for $|M| = 0$ the spectrum of $H_{UTe_2}$ was solved exactly. The quasiparticle eigenstates $E(k_x = 0, k_y)$ versus $k_y$ have been predicted for the chiral, time reversal symmetry breaking, *p*-wave order parameter with *Au* + *iB3u* symmetry. Here, a chiral QSB spans the full energy range $-\Delta_{UTe_2} \leq E \leq \Delta_{UTe_2}$, crossing the Fermi level ($E = 0$) and generating a finite density of quasiparticle states $N(|E| < \Delta_{UTe_2})$. The quasiparticle spectrum versus $k_y$ at $k_x = 0$ were predicted for non-chiral, time reversal symmetry conserving, *p*-wave order parameter with B3u symmetry. Here, two non-chiral QSBs also span the full energy range $-\Delta_{UTe_2} \leq E \leq \Delta_{UTe_2}$, and feature $E = 0$ states, thus generating a finite $N(|E| < \Delta_{UTe_2})$. Although these QSBs have dispersion in both the positive and negative $k_y$ directions and can backscatter, their gaplessness is protected by time reversal symmetry with $T^2 = -I$.

To distinguish a chiral from non-chiral $\Delta_\mathbf{k}$ by using SATM within the SIP model requires quantitative calculation of the Andreev conductance $a(V) = dI/dV|_{SIP}$ between Nb and UTe2 using the non-chiral QSB to demonstrate that a sharp $a(V)$ peak should occur surrounding zero-bias [65]. Because Andreev reflection of QSB quasiparticles allows highly efficient transfer of charge 2*e* across the junction, its sharpness is robust meaning that Andreev transport between *s*-wave/*p*-wave electrodes through a QSB, makes scanned Andreev tunneling spectroscopy an ideal new approach for studying superconductive topological quasiparticle surface bands of ITS. In the limit where the tunneling matrix element to the *s*-wave electrode $|M| \to 0$ these phenomena are indistinguishable but, as $|M|$ increases, the wavefunctions of the Nb overlap those of UTe2 allowing detection of the QSB quasiparticles at the *s*-wave electrode. The predicted quasiparticle bands within the SIP interface between Nb and UTe₂ for the chiral order parameter $A_u + iB_{3u}$ symmetry as a function of increasing $|M|$ were predicted. With increasing $|M|\sim 1/R$ where $R$ is the SIP tunnel junction resistance, the proximity effect of the $s$-wave electrode generates two chiral QSBs for all $|E| < \Delta_{\text{UTe}_2}$, both of which cross $E = 0$. Hence, for the chiral $\Delta_k$, the zero-energy $N(E)$ will be virtually unperturbed by increasing $|M|$. Likewise, the QSB within the SIP interface as a function of $|M|$ for the non-chiral order parameter with $B_{3u}$ symmetry were also predicted. When $|M| \to 0$ the non-chiral QSB crosses $E = 0$. But, with increasing $|M|\sim 1/R$, time reversal symmetry breaking due to the $s$-wave electrode splits the QSB of quasiparticle into two, neither of which cross $E = 0$. This reveals that the zero-energy $N(E)$ peak must split as the zero-energy quasiparticles of the QSB disappear, generating two particle-hole symmetric $N(E)$ maxima at finite energy. The $N(0)$ is quantitatively predicted to split into two particle-hole symmetric $N(E)$ maxima as a function of $|M|$ for a chiral $\Delta_k$ but not for a non-chiral $\Delta_k$. Thus, in theory, Andreev tunneling between an $s$-wave electrode and a $p$-wave superconductor through the latter's QSB, is that a non-chiral pairing state can be distinguished from a chiral pairing state [65].

Modeling the QPI signature of the QSB was the next challenge. Here it is the normal state electronic structure of UTe₂ forms the basis upon which $\Delta(\boldsymbol{k})$ phenomenology emerges at lower temperatures. Atomic-resolution differential tunneling conductance $g(\boldsymbol{r}, V) \equiv dI/dV(\boldsymbol{r}, V)$ imaging visualizes the density-of-states $N(\boldsymbol{r}, E)$ and its Fourier transform $g(\boldsymbol{q}, E) \propto N(\boldsymbol{q}, E)$ can be used to establish electronic-structure characteristics. Hence, a conventional model of the bulk first BZ of UTe₂ sustaining a two-band Fermi surface (FS) as now widely hypothesized [66,67]. Quantitative predictions for the normal state QPI in UTe₂ then require a Hamiltonian $H_{\text{UTe}_2} = \begin{pmatrix} H_{U-U} & H_{U-Te} \\ H_{U-Te}^+ & H_{Te-Te} \end{pmatrix}$ such that $H_{U-U}$ and $H_{Te-Te}$ describe respectively the two uranium and tellurium orbitals and $H_{U-Te}$ their hybridization. From this one anticipates strong scattering interference with a sextet of wavevectors $\boldsymbol{p}_i: i = 1 - 6$ viewed from the (001) plane, where $a$ is the $x$-axis unit cell distance and $b$ is the $y$-axis unit cell distance.

| Wavevector | $\boldsymbol{p}_1$ | $\boldsymbol{p}_2$ | $\boldsymbol{p}_3$ | $\boldsymbol{p}_4$ | $\boldsymbol{p}_5$ | $\boldsymbol{p}_6$ |
|---|---|---|---|---|---|---|
| Coordinate $(\frac{2\pi}{a}, \frac{2\pi}{b})$ | (0.29,0) | (0.43, 1) | (0.29,2) | (0, 2) | (−0.14, 1) | (0.57,0) |

However, the natural cleave surface of UTe₂ crystal is not (001) but rather (0-11), here shown schematically in Fig. 2a, and it is this surface that the scan-tip approaches perpendicularly. To clarify the normal state band structure and quasiparticle interference viewed from (0-11) plane, the $\boldsymbol{k}$-space joint density of states $J(\boldsymbol{q}, E)$ was calculated at the (001) plane using the UTe₂ FS that takes into account the uranium $f$ orbital spectral weight. The sextet of scattering wavevectors $\boldsymbol{p}_i: i = 1 - 6$ derived heuristically above are then revealed as primary peaks in $J(\boldsymbol{q}, E)$. Here, $J(\boldsymbol{q}, E)$ for the same band-structure model has been calculated but viewed along the normal to the (0-11) plane [68]. Here the $y$-coordinates of the (0-11) sextet become $\boldsymbol{q}_{1,y} = \boldsymbol{p}_{1,y}\sin\theta$ where $\theta = 24°$ and $c^*$ is the (0-11) surface $y:z$-axis lattice periodicity, as indicated by the colored arrows in Ref. [68].

| Wavevector | $\boldsymbol{q}_1$ | $\boldsymbol{q}_2$ | $\boldsymbol{q}_3$ | $\boldsymbol{q}_4$ | $\boldsymbol{q}_5$ | $\boldsymbol{q}_6$ |
|---|---|---|---|---|---|---|
| Coordinate $(\frac{2\pi}{a}, \frac{2\pi}{c^*})$ | (0.29,0) | (0.43, 0.5) | (0.29,1) | (0, 1) | (−0.14, 0.5) | (0.57,0) |

In UTe₂, the $A_u$ state should be completely gapped on both Fermi surfaces whereas $B_{1u}$, $B_{2u}$ and $B_{3u}$ states could exhibit point nodes along the $k_z$ −axis, $k_y$ −axis and $k_x$ −axis, respectively. These bulk Bogoliubov eigenstates are described by the dispersion

$$E_k = \sqrt{\xi_k^2 + \Delta^2(|\boldsymbol{d}(\boldsymbol{k})|^2 \pm |\boldsymbol{d}(\boldsymbol{k}) \times \boldsymbol{d}^*(\boldsymbol{k})|)} \tag{2}$$

so that $\boldsymbol{k}$-space locations of energy-gap zeros are defined in general by $|\boldsymbol{d}(\boldsymbol{k})|^2 \pm |\boldsymbol{d}(\boldsymbol{k}) \times \boldsymbol{d}^*(\boldsymbol{k})| = 0$. Thus, although $A_u$ supports no energy-gap nodes by definition and $B_{1u}$ exhibits no energy-gap nodes in this model, there are numerous nodes in highly distinct $\boldsymbol{k}$-space nodal locations for $B_{2u}$ and $B_{3u}$. The bulk FSs have energy-gap nodal locations for $B_{1u}$, $B_{2u}$ and $B_{3u}$ from Eqn. 4. QPI predictions for the QSB in UTe₂ used the Hamiltonian

$$H(k) = \begin{pmatrix} H_{UTe_2}(k) \otimes I_2 & \Delta(k) \otimes I_4 \\ \Delta^+(k) \otimes I_4 & -H^*_{UTe_2}(-k) \otimes I_2 \end{pmatrix} \tag{3}$$

where the order parameter is $\Delta(k) = \Delta_0(\boldsymbol{d} \cdot \boldsymbol{\sigma})i\sigma_2$ and $I_2$, $I_4$ are the unit matrices. The focus primarily was on $B_{2u}$ and $B_{3u}$:

$$\boldsymbol{d}_{B_{2u}} = \big(C_1 sin(k_z c), C_0 sin(k_x a) sin(k_y b) sin(k_z c), C_3 sin(k_x a)\big) \tag{4a}$$

$$\boldsymbol{d}_{B_{3u}} = \big(C_0 sin(k_x a) sin(k_y b) sin(k_z c), C_2 sin(k_z c), C_3 sin(k_y b)\big) \tag{4b}$$

where $a, b, c$ are lattice constants, and $C_0 = 0$, $C_1 = 300$ µeV, $C_2 = 300$ µeV, and $C_3 = 300$ µeV. The unperturbed bulk Green's function is then: $G_0(\boldsymbol{k}, E) = [(E + i\eta)I - H(\boldsymbol{k})]^{-1}$ ($\eta = 100$ µeV) with the corresponding unperturbed spectral function: $A_0(\boldsymbol{k}, E) = -1/\pi \ Im \ G_0(\boldsymbol{k}, E)$. The surface Green's function $G_s(\boldsymbol{k}, E)$ characterizes a semi-infinite system with broken translation symmetry and therefore cannot be calculated directly. A novel technique was used to model the surface using a strong planar impurity [69-71]. In the limit of an infinite impurity potential, the impurity plane splits the system into two semi-infinite spaces. So the only wavevectors in the (0-11) plane remain good quantum numbers. The effect of the planar-impurity can then be exactly calculated using the T-matrix formalism which gives one access to the surface Green's function of the semi-infinite system. For Bogoliubov QPI predictions at the (0-11) surface of UTe₂, a localized impurity potential $\hat{V} = V\tau_z \otimes I_8$ where $V = 0.2$ eV was used to determine the surface Green's function $g_S(\boldsymbol{q}, \boldsymbol{k}, E)$ using the T-matrix $T(E) = \left(I - \hat{V} \int \frac{d^2k}{S_{BZ}} G_s(\boldsymbol{k}, E)\right)^{-1} \hat{V}$. Then the QPI patterns for the UTe₂ QSB are predicted directly using:

$$N(\boldsymbol{q}, E) = \frac{i}{2\pi} \int \frac{d^2k}{S_{BZ}} Tr[g_S(\boldsymbol{q}, \boldsymbol{k}, E)] \tag{5}$$

where

$$g_S(\boldsymbol{q}, \boldsymbol{k}, E) = G_S(\boldsymbol{q}, E)T(E)G_S(\boldsymbol{q} - \boldsymbol{k}, E) - G_S^*(\boldsymbol{q} - \boldsymbol{k}, E)T^*(E)G_S^*(\boldsymbol{q}, E) \tag{6}$$

**SATM Experiments on UTe₂**

To explore UTe₂ for such conjectured ITS phenomenology, single-crystal samples are introduced to a superconductive-tip scanning tunneling microscope [53-58], cleaved at 4.2 K in cryogenic ultrahigh vacuum, inserted to the scan head, and cooled to $T = 280$ mK. A typical topographic image $T(\boldsymbol{r})$ of the (0-11) cleave surface as measured by a superconductive Nb tip is shown in Fig. 2b with atomic periodicities defined by vectors $\boldsymbol{a^*}$, $\boldsymbol{b^*}$, where $\boldsymbol{a^*}=\boldsymbol{a}= 4.16$ Å is the $\hat{x}$-axis unit-cell vector and $\boldsymbol{b^*}= 7.62$ Å is a vector in the $\hat{y}:\hat{z}$ plane. As the temperature is reduced, a sharp zero-energy peak appears within the overall energy gap in the spectrum (Fig. 3a). This robust zero-bias $dI/dV|_{\text{SIP}}$ peak is observed universally, as exemplified for example by Figs. 3b, c. One sees that these phenomena are not due to Josephson tunneling because the zero-bias conductance $a(0)$ of Nb/UTe$_2$ is many orders of magnitude larger than it could possibly be due to Josephson currents through the same junction [65], and because $a(0)$ grows linearly with falling $R$ before diminishing steeply as $R$ is further reduced while $g(0)$ due to Josephson currents grow continuously as $1/R^2$. Moreover, the SIP model predicts quantitatively that such an intense $a(0)$ peak should occur if $\Delta_{\boldsymbol{k}}$ of UTe$_2$ supports a QSB within the interface (Fig. 1) and because Andreev transport due to these QSB quasiparticles allows a strong zero-bias conductance to the Nb electrode.

To determine spectroscopically whether the UTe$_2$ order parameter is chiral, the evolution of Andreev conductance $a(V)$ at $T$ = 280 mK was measured as a function of decreasing junction resistance $R$ or equivalently increasing tunneling matrix element $|M|$. Figure 4a shows the strong energy splitting $\delta E$ observable in $a(V)$, that first appears and then evolves with increasing $1/R$. Figure 4b shows the measured $a(\boldsymbol{r}, V)$ splitting across the (0 -1 1) surface of UTe$_2$ along the arrow indicated in Fig. 3b, demonstrating that $a(\boldsymbol{r}, V)$ split-peaks are pervasive. Decisively, we plot in Fig. 4c the measured $\delta E$ between peaks in $a(\boldsymbol{r}, V)$ at $T$ = 280 mK versus $1/R$. On the basis of predictions for energy splitting $\delta E$ within the SIP model [65] for chiral $\Delta_{\boldsymbol{k}}$ and non-chiral $\Delta_{\boldsymbol{k}}$, the chiral $\Delta_{\boldsymbol{k}}$ appears ruled out.

For QSB QPI studies, Figure 5a shows a typical 66 nm square field-of-view (FOV) topography of the (0-11) cleave surface which can be studied both in the normal and superconducting states. Figure 5b shows typical $dI/dV$ spectra measured with a superconductive tip in both the normal state at 4.2 K and the superconducting state at 280 mK, far below $T_C$. In the latter case, two intense joint-coherence peaks are located at $E = \Delta_{\text{Nb}} + \Delta_{\text{UTe}_2}$. More importantly, a high density of QSB quasiparticles allows efficient creation and annihilation of Cooper pairs in both superconductors, thus generating intense Andreev differential conductance $a(\boldsymbol{r}, V) \equiv dI/dV|_A (\boldsymbol{r}, V)$ for $|V| < \Delta_{\text{UTe}_2}/e \sim 300$ μV as indicated by yellow shading. Compared to conventional NIS tunneling using a normal metallic tip, this Andreev conductance provides a significant improvement in the energy resolution ($\delta E \sim 10$ μeV) of QSB scattering interference measurements. Comparing measured $g(\boldsymbol{r}, V):g(\boldsymbol{q}, V)$ recorded in the normal state at 4.2 K (Fig. 5c) with measured $a(\boldsymbol{r}, V): a(\boldsymbol{q}, V)$ in the superconducting state at 280 mK (Fig. 5d), both with identical FOV and junction characteristics, allows determination of which phenomena at the (0-11) surface emerge only due to superconductivity. Several peaks of the sextet are present in the normal state $g(\boldsymbol{q}, V)$ in Fig. 5c as they originate from scattering of the normal state band structure [68]. The complete predicted QPI sextet $\boldsymbol{q}_i: i = 1 - 6$ are only detected in the superconducting state and appears to rely on scattering between QSB states. The sextet wavevectors are highlighted by colored arrows in Fig. 5d. The experimental maxima in $a(\boldsymbol{q}, V)$ and the theoretically predicted $\boldsymbol{q}_i$ from Ref. [68], are in excellent quantitative agreement with a maximum 3% difference between all their wavevectors. This demonstrated, for the first time, that the FS which dominates the bulk electronic structure of UTe$_2$ is also what controls QSB $\boldsymbol{k}$-space geometry at its cleave surface. Furthermore, Fig. 5e reveals how the amplitudes of the superconducting state QPI are enhanced compared to the normal state measurements. The predominant effects of bulk superconductivity are the strongly enhanced arc-like scattering intensity connecting $q = 0$ and $q_5$ and the unique appearance of wavevector $q_1$.

To visualize the QSB dispersion $k(E)$ of UTe$_2$ we next use superconductive-tip $a(r,V)$: $a(q,V)$ measurements to image energy resolved QPI at the (0-11) cleave surface. Figure 6a presents the measured $a(r,V)$ at $|V| = 0\ \mu V, 50\ \mu V, 100\ \mu V, 150\ \mu V, 200\ \mu V, 250\ \mu V$ recorded at $T$ = 280 mK in the identical FOV as Fig. 5a. These data are highly typical of such experiments in UTe$_2$ [68]. Figure 6b contains the consequent scattering interference patterns $a(q,V)$ at $|V| = 0\ \mu V, 50\ \mu V, 100\ \mu V, 150\ \mu V, 200\ \mu V, 250\ \mu V$ as derived by Fourier analysis of Fig. 6a. Here the energy evolution of scattering interference of the QSB states is obvious. For comparison with theory, detailed predicted characteristics of $N(q,E)$ for a $B_{2u}$-QSB and $B_{3u}$-QSB at the (0-11) SBZ were determined in Ref. [68]; here again energies range $|E| = 0\ \mu eV, 50\ \mu eV, 100\ \mu eV, 150\ \mu eV, 200\ \mu eV, 250\ \mu eV$. Each QPI wavevector is determined by maxima in the $N(q,E)$ QPI pattern (Fig. 6b); these phenomena are highly repeatable in multiple independent experiments. The strongly enhanced QPI features occurring along the arc connecting $q = 0$ and $q_5$ (Fig. 6b) are characteristic of the theory for a $B_{3u}$-QSB [68]. Most critically, the intense QPI appearing at wavevector $q_1$ (yellow circle in Fig. 6b) is a characteristic of the $B_{3u}$ superconducting state, deriving from its geometrically unique nodal structure [68]. The appearance of scattering interference of QSB quasiparticles at $q_1$ in the superconducting state (Fig. 5d and 6b) is precisely as would be anticipated in theory [51,52] due to projection of $B_{3u}$ energy-gap nodes on the bulk FS [68] onto the (0-11) crystal surface 2D Brillouin zone.

**CONCLUSIONS**

Overall, the chiral order parameters $A_u + iB_{1u}$ and $B_{3u} + iB_{2u}$ proposed for UTe$_2$ appear inappropriate because of the observed Andreev conductance $a(0)$ splitting when reducing the Nb/UTe$_2$ separation [65]. Within the four possible odd-parity $\Delta_k$ symmetries $A_u$, $B_{1u}$, $B_{2u}$ and $B_{3u}$, the isotropic $A_u$ order parameter also appears insupportable because its QSB is a Majorana-cone of Bogoliubons with zero density-of-states at zero energy meaning that Andreev conductance $a(0)$ would be highly suppressed. Andreev conductance between Nb (*s*-wave) and UTe$_2$ (putative *p*-wave) superconductors allows visualization of a powerful zero-energy $a(V) = dI/dV|_\text{SIP}$ peak at the UTe$_2$ (0-11) surface. And, with enhanced tunneling to an *s*-wave electrode (Nb) this zero-energy Andreev spectrum splits strongly into two finite-energy conductance maxima [65]. Moreover, visualizing dispersive QSB scattering interference reveals unique in-gap QPI patterns exhibiting a characteristic sextet of wavevectors $q_i$: $i = 1 - 6$ due to projection of the bulk superconductive band structure onto the (0-11) surface [68]. Although $q_2$ and $q_6$ are weakly observable in the normal state, features at $q_5$ and $q_6$ become strongly enhanced for superconducting state QPI at $|E| < \Delta$ and QPI appears at wavevector $q_1$ uniquely in the superconducting state. This complete phenomenology, by correspondence with theory [65,68], is most consistent with a 3D, odd-parity, spin-triplet, time-reversal-symmetry conserving, *a*-axis nodal superconducting order parameter with $B_{3u}$ symmetry in UTe$_2$.

Figure 1

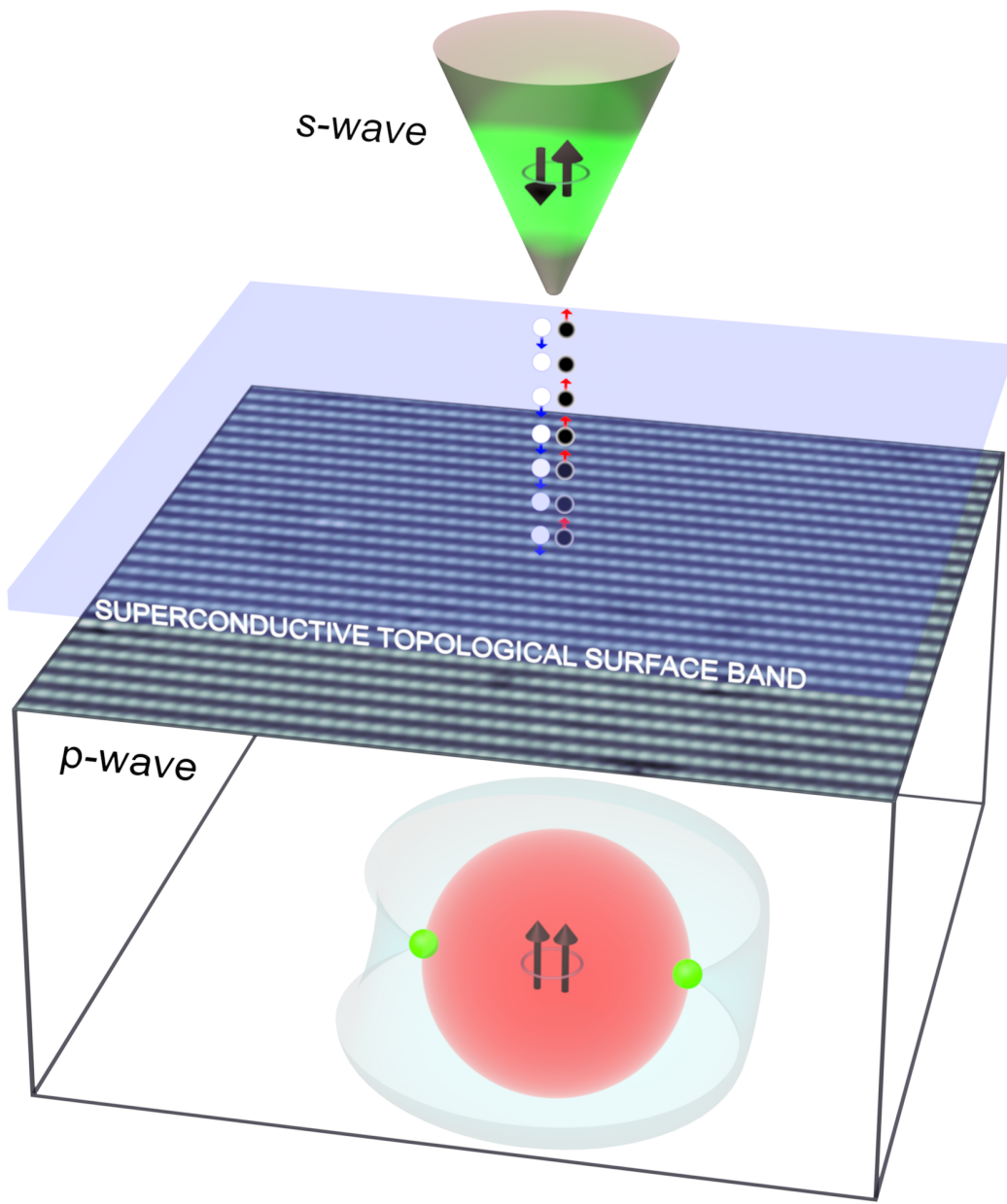

**FIG. 1**

Schematic of SIP (*s*-wave to *p*-wave) tunneling through a superconductive topological surface state. This is the basic technique implemented throughout this paper.

Figure 2

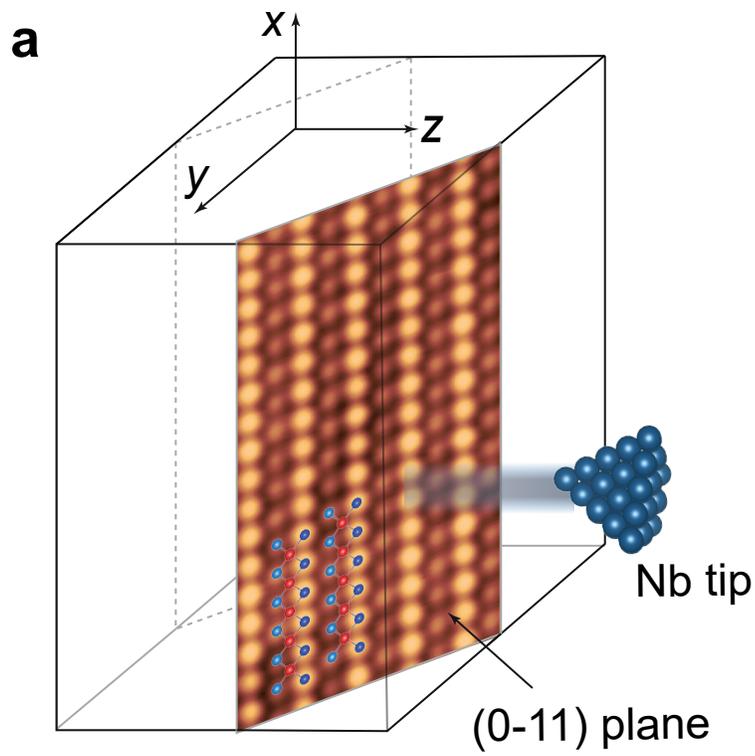

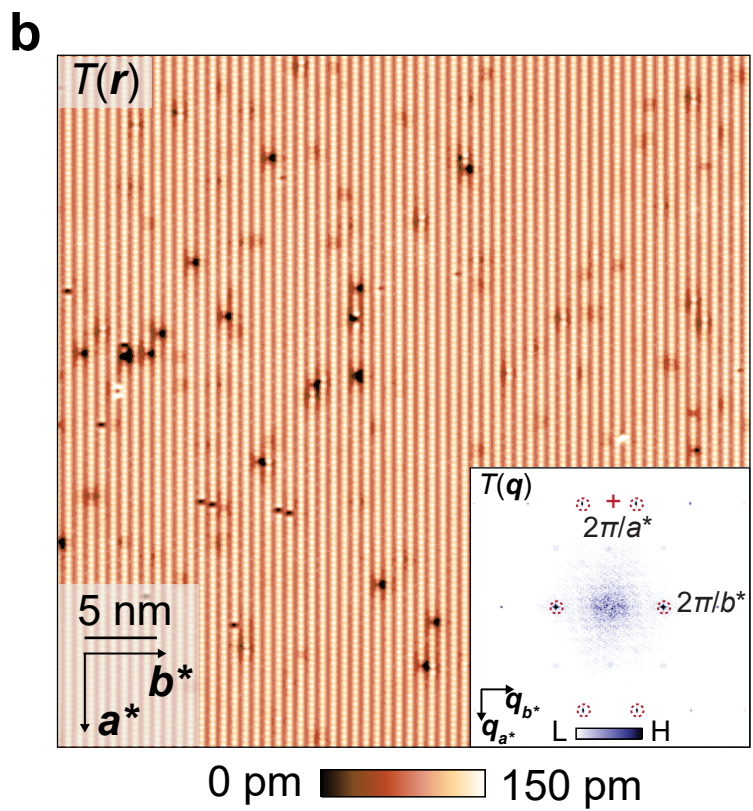

**Fig. 2**

a. Schematic of (0-11) cleave surface of $UTe_2$ shown in relative orientation to the STM tip tunneling direction. Measured high-resolution $T(r)$ at low junction resistance ($I_s$ = 3 nA, $V_s$ = 5 mV), clarifying two types of Te atom in light and dark blue. The U atoms are indicated in red.
b. Typical topographic image $T(r)$ of $UTe_2$ (0–11) surface measured with a superconducting tip at $T$ = 280 mK ($I_s$ = 0.5 nA, $V_s$ = 30 mV). Inset: measured $T(q)$, the Fourier transform of $T(r)$ in **b**, with the surface reciprocal-lattice points labelled as dashed red circles.

Figure 3

**a**

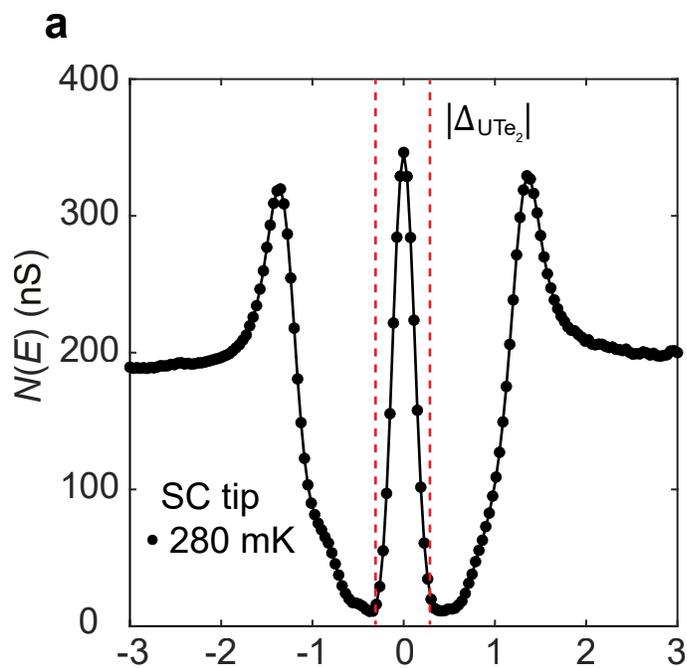

**b**

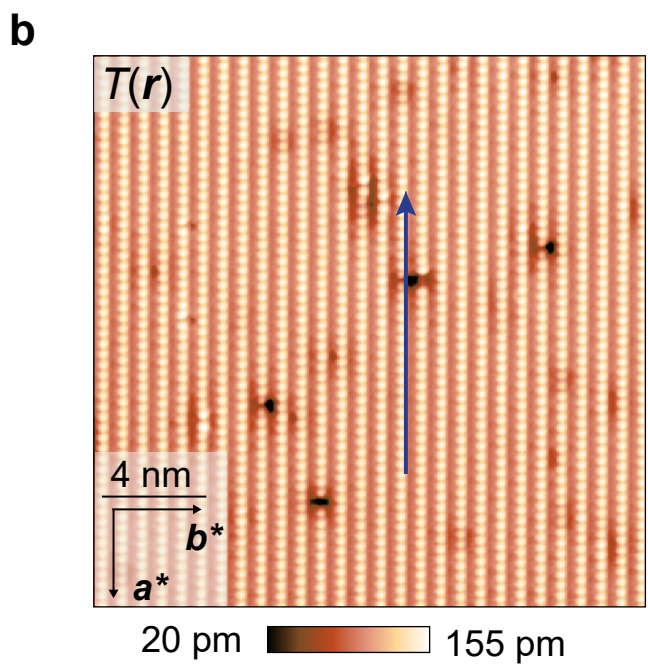

**c**

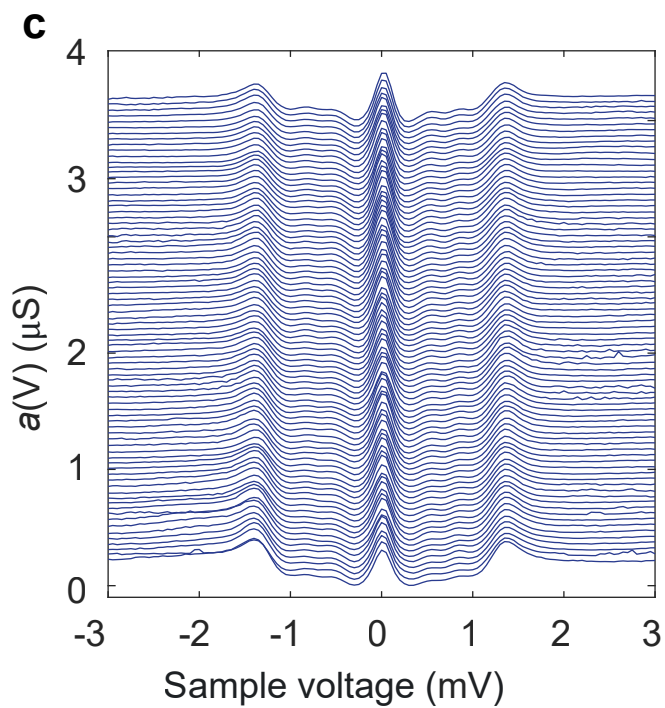

**Fig. 3**

a. Typical SIP Andreev conductance spectrum $a(V) \equiv dI/dV|_{\text{SIP}}$ measured with Nb scan-tip on UTe$_2$ (0-11) surface for junction resistance $R$ = 6 MΩ and $T$ = 280 mK. A high intensity zero-bias $dI/dV|_{\text{SIP}}$ peak is observed.
b. Typical topographic image $T(r)$ of (0-11) surface ($I_s$ = 0.2 nA, $V_s$ = 5 mV).
c. Evolution of measured $a(r,V)$ across the (0-11) surface of UTe$_2$ indicated by the arrow in Fig. 3b for junction resistance $R$ = 6 MΩ and $T$ = 280 mK. The zero-bias $dI/dV|_{\text{SIP}}$ peaks are universal and robust, indicating that the zero energy ABS is omnipresent.

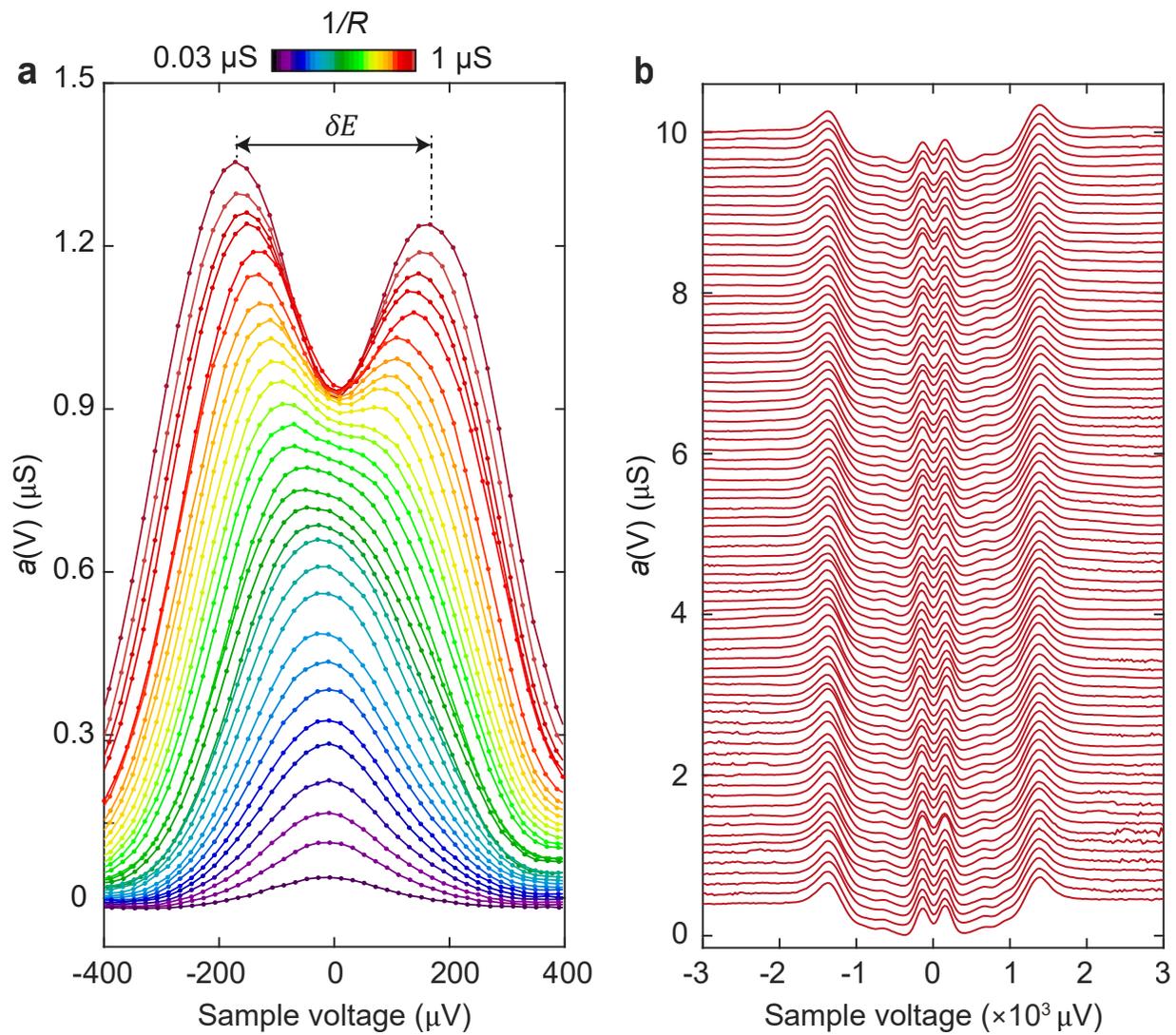

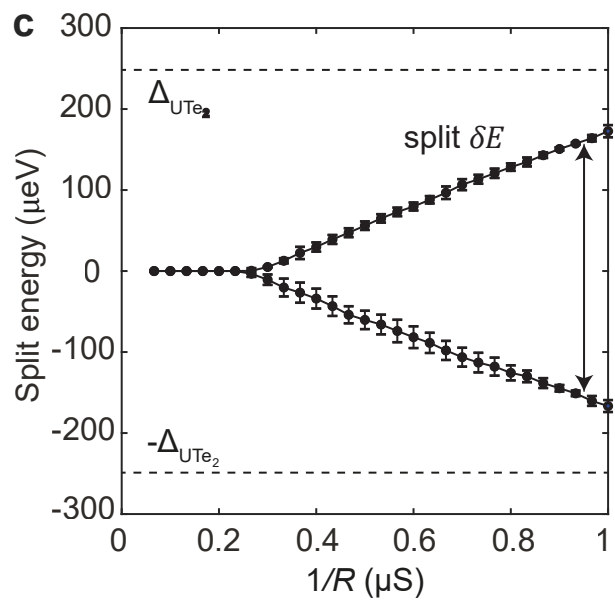

**Fig. 4**

a. Measured evolution of $a(V) \equiv dI/dV|_{\text{SIP}}$ at $T$ = 280 mK in UTe$_2$ as a function of decreasing junction resistance $R$ and thus increasing tunneling matrix element $|M| \sim 1/R$. When the junction resistance falls below $R \sim 5$ MΩ, the $a(V)$ spectra start to split.
b. Evolution of measured $a(\mathbf{r},V)$ splitting across the (0 -1 1) surface of UTe$_2$ at junction resistance $R$ = 3 MΩ and $T$ = 280 mK, demonstrating that $a(\mathbf{r},V)$ split-peaks are pervasive at low junction resistance $R$ and high tunneling matrix $|M|$.
c. Measured energy splitting of $a(0)$ at $T$ = 280 mK in UTe$_2$ versus $1/R$. These data may be compared with predictions of $a(V)$ splitting for $A_u + iB_{3u}$ and $B_{3u}$ order parameters [65] of UTe$_2$.

Figure 5

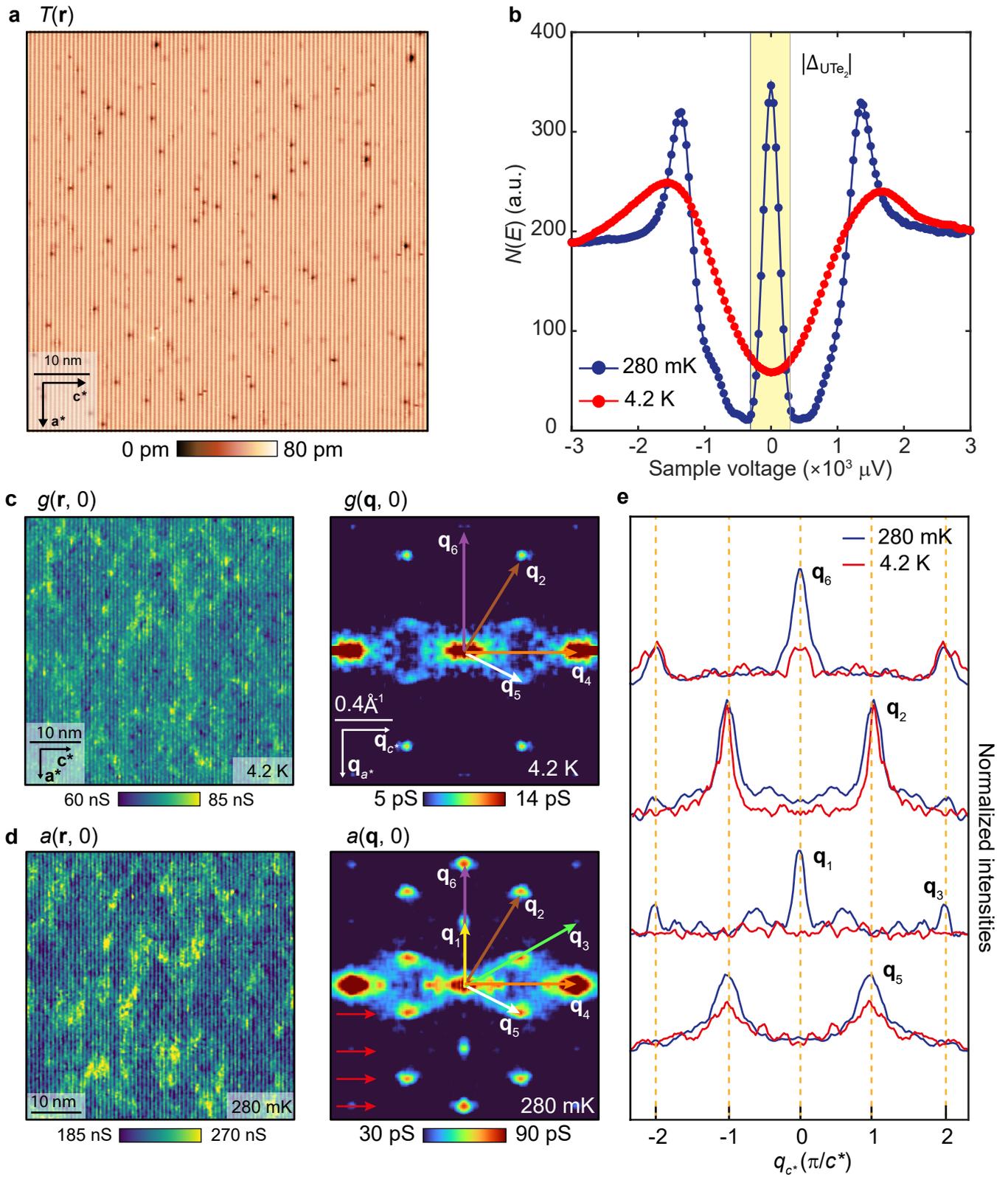

**Fig. 5**
a. Typical topographic image $T(r)$ of the (0-11) cleave surface of UTe$_2$ where QPI patterns are imaged.
b. Measured differential conductance in the UTe$_2$ normal state $g(V)$ at $T$=4.2 K (red curve); and Andreev differential conductance in the superconducting state $a(V)$ at $T$=280 mK (blue curve). Intense Andreev conductance is observed at $V = 0$.
c. Measured $g(r, 0)$ and $g(q, 0)$ at $T$=4.2 K in the UTe$_2$ normal state in the identical FOV as **a**. The setpoint is $V_s = 3$ mV and $I = 200$ pA.
d. Measured $a(r, 0)$ and $a(q, 0)$ at $T$=280 mK in the UTe$_2$ superconducting state in the identical FOV as **a** and **c**. Here a sextet of scattering interference wavevectors $q_i$, $i$ = 1-6 are identified. This experimental detection of the sextet has been repeated multiple times [68]. The set point is $V_s = 3$ mV and $I = 200$ pA.
e. Relative amplitudes of the sextet wavevectors in the normal and superconducting states. Comparison of $g(q, 0)$ linecuts at $T$=4.2 K and $a(q, 0)$ linecuts measured $T = 280$ mK. The linecuts are taken horizontally in the $q$ space indicated by red arrow in **d**. The linecuts have been normalized by their background intensities at 280 mK and 4.2 K. The intensities of $q_5$ and $q_6$ are significantly enhanced in the superconducting state. Most importantly, $q_1$ only appears in the superconducting state.

Figure 6

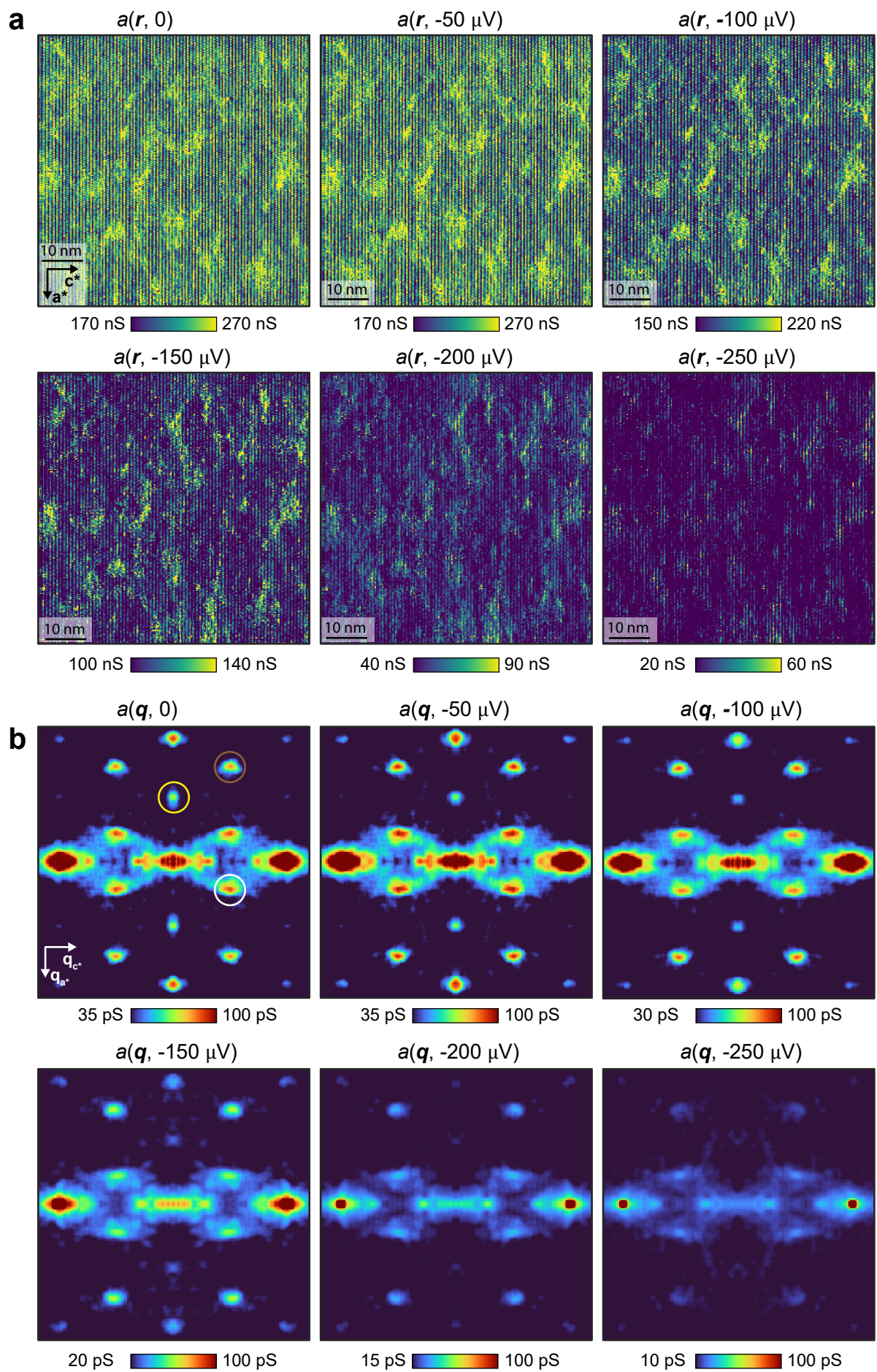

**Fig. 6**

a. Measured $a(r, V)$ at the (0-11) cleave plane of UTe$_2$ at bias voltages $|V|$ = 0 μV, 50 μV, 100 μV, 150 μV, 200 μV, 250 μV. The setpoint is $V_s$ = 3 mV and $I$ = 200 pA.

b. Measured $a(q, V)$ at the (0-11) cleave plane of UTe$_2$ at bias voltages $|V|$ = 0 μV, 50 μV, 100 μV, 150 μV, 200 μV, 250 μV. The setpoint is $V_s$ = 3 mV and $I$ = 200 pA. Each QPI wavevector in this FOV, $q_1$ (yellow), $q_2$ (brown) and $q_5$ (white), is identified as the maxima position (colored circles) in the QPI data. Particularly $q_1$ is a characteristic only of the $B_{3u}$ superconducting state [68] and it only exists inside the energy gap.

**Author Contributions**      J.C.S.D. and S.W. wrote the paper.

**Acknowledgments**      We acknowledge and very gratefully thank all the collaborators who carried out this research campaign: Qiangqiang Gu, Joseph P. Carroll, Kuanysh Zhussupbekov, Bin Hu, Xiaolong Liu, Dung-Hai Lee, Catherine Pepin, Cristina Bena, Adeline Crépieux, Emile Pangburn, Sheng Ran, Christopher Broyles and Johnpierre Paglione. S.W. and J.C.S.D. acknowledge support from the European Research Council (ERC) under Award DLV-788932 and the Moore Foundation's EPiQS Initiative through Grant GBMF9457. S.W. acknowledges support from the Engineering and Physical Sciences Research Council (EPSRC) under Award EP/Z53660X/1 and the support from the Royal Academy of Engineering / Leverhulme Trust Research Fellowship. J.C.S.D. acknowledges support from the Royal Society under Award R64897 and Science Foundation Ireland under Award SFI 17/RP/5445.

**Data Availability**   No datasets were generated or analyzed during the current study.

**Competing interests**   The authors declare no competing interests.

Parameter in UTe$_2$ Determined by Knight Shift Measurement, *J. Phys. Soc. Jpn.* **91**, 043705 (2022).

17. T. Metz, S. Bae, S. Ran, I-L. Liu, Y. S. Eo, W. T. Fuhrman, D. F. Agterberg, S. M. Anlage, N. P. Butch, J. Paglione, Point-node gap structure of the spin-triplet superconductor UTe$_2$, *Phys. Rev. B* **100**, 220504(R) (2019).
18. S. Suetsugu, M. Shimomura, M. Kamimura, T. Asaba, H. Asaeda, Y. Kosuge, Y. Sekino, S. Ikemori, Y. Kasahara, Y. Kohsaka, M. Lee, Y. Yanase, H. Sakai, P. Opletal, Y. Tokiwa, Y. Haga, Y. Matsuda, Fully gapped pairing state in spin-triplet superconductor UTe$_2$, *Sci. Adv.* **10**, 2375-2548 (2024).
19. S. Kittaka, Y. Shimizu, T. Sakakibara, A. Nakamura, D. Li, Y. Homma, F. Honda, D. Aoki, K. Machida, Orientation of point nodes and nonunitary triplet pairing tuned by the easy-axis magnetization in UTe$_2$, *Phys. Rev. Res.* **2**, 032014(R) (2020).
20. I. M. Hayes, D. S. Wei, T. Metz, J. Zhang, Y. S. Eo, S. Ran, S. R. Saha, J. Collini, N. P. Butch, D. F. Agterberg, A. Kapitulnik, J. Paglione, Multicomponent superconducting order parameter in UTe$_2$, *Science* **373**, 797–801 (2021).
21. L. P. Cairns, C. R. Stevens, C. D. O'Neill, A. Huxley, Composition dependence of the superconducting properties of UTe$_2$, *J. Phys.: Condens. Matter* **32**, 415602 (2020).
22. Y. Iguchi, H. Man, S. M. Thomas, F. Ronning, P.F. S. Rosa, K. A. Moler, Microscopic Imaging Homogeneous and Single Phase Superfluid Density in UTe$_2$, *Phys. Rev. Lett.* **130**, 196003 (2023).
23. K. Ishihara, M. Roppongi, M. Kobayashi, K. Imamura, Y. Mizukami, H. Sakai, P. Opletal, Y. Tokiwa, Y. Haga, K. Hashimoto, T. Shibauchi, Chiral superconductivity in UTe$_2$ probed by anisotropic low-energy excitations, *Nat. Commun.* **14**, 2966 (2023).
24. L. Jiao, S. Howard, S. Ran, Z. Wang, J. O. Rodriguez, M. Sigrist, Z. Wang, N. P. Butch, V. Madhavan, Chiral superconductivity in heavy-fermion metal UTe$_2$, *Nature* **579**, 523–527 (2020).
25. M. O. Ajeesh, M. Bordelon, C. Girod, S. Mishra, F. Ronning, E. D. Bauer, B. Maiorov, J. D. Thompson, P. F. S. Rosa, S. M. Thomas, Fate of time-reversal symmetry breaking in UTe$_2$, *Phys. Rev. X* **13**, 041019 (2023).
26. Q.-H. Wang and D.-H. Lee, Quasiparticle scattering interference in high-temperature superconductors, *Phys. Rev. B* **67**, 020511 (2003).
27. L. Capriotti, D. J. Scalapino, and R. D. Sedgewick, Wave-vector power spectrum of the local tunneling density of states: Ripples in a d-wave sea, *Phys. Rev. B* **68**, 014508 (2003).
28. J.S. Hofmann, R. Queiroz, A.P. Schnyder, Theory of quasiparticle scattering interference on the surface of topological superconductors, *Phys. Rev. B* **88**, 134505 (2013).
29. J. E. Hoffman et al. Imaging Quasiparticle Interference in Bi$_2$Sr$_2$CaCu$_2$O$_{8+\delta}$, *Science* **297**, 5584 1148-1151 (2002).*24*